\documentclass[prd,aps,showpacs,tightenlines,superscriptaddress,amsmath,amssymb,amsfonts,twocolumn]{revtex4-2}
\usepackage{amssymb,amsmath,amsthm,amsbsy,epsfig,color,graphicx,times}
\usepackage[ansinew]{inputenc}
\usepackage[english]{babel}
\usepackage{color}
\usepackage{braket}
\usepackage{tabularx}
\usepackage{array}
\usepackage{slashed}

\begin{document}

\title{Muon $g-2$ anomaly and non-locality}

\author{A. Capolupo}
\email{capolupo@sa.infn.it}
\affiliation{Dipartimento di Fisica ``E.R. Caianiello'' Universit\`{a} di Salerno, and INFN --- Gruppo Collegato di Salerno, Via Giovanni Paolo II, 132, 84084 Fisciano (SA), Italy}

\author{G. Lambiase}
\email{lambiase@sa.infn.it}
\affiliation{Dipartimento di Fisica ``E.R. Caianiello'' Universit\`{a} di Salerno, and INFN --- Gruppo Collegato di Salerno, Via Giovanni Paolo II, 132, 84084 Fisciano (SA), Italy}

\author{A. Quaranta}
\email{anquaranta@unisa.it}
\affiliation{Dipartimento di Fisica ``E.R. Caianiello'' Universit\`{a} di Salerno, and INFN --- Gruppo Collegato di Salerno, Via Giovanni Paolo II, 132, 84084 Fisciano (SA), Italy}

\begin{abstract}

We show that the discrepancy between the observed value of the muon anomalous moment and the standard model prediction
can be explained in the framework of nonlocal theories.
We compute the leading order and next to leading order nonlocal correction to the anomalous magnetic moment $  \alpha_{NL}$ and we find that  it depends   on the nonlocality scale $M_f$ and the fermion mass $m_f$ as $\alpha_{NL} \propto \frac{m_f^2}{M_f^2}$.
Such a dependence of the anomalous magnetic momentum allows to explain, in a flavor-blind nonlocality scale, why the observed anomalous magnetic moment of the electron is  much closer to the standard model prediction, and permits to predict  a large anomaly that should exist for the $\tau$ particle. We also determine the lower bounds on the nonlocality scale, for both flavor-blind and flavor-dependent scenarios. 

\end{abstract}

\maketitle

\section{Introduction}

In recent years many phenomena, ranging from the strong CP problem, particle mixing and oscillations, matter-antimatter asymmetry, to the nature of dark matter and dark energy, as well as the lack of a quantum theory of gravity, have lead to the development of theories beyond the Standard Model (SM) of particles \cite{BSM1,BSM2,BSM3,BSM4,BSM5,BSM6}. An example of a phenomenon which cannot be described within the SM may be represented by the muon magnetic moment anomaly.

The successful prediction of the magnetic moment of the charged leptons is one of the most celebrated accomplishments in Quantum Field Theory. The Dirac theory predicts a $g$-factor $g_{Dirac} = 2$, whereas the radiative corrections shift the actual value slightly above $2$. This allows one to define the anomalous magnetic moment $\alpha$ as the difference between the quantum field theoretic prediction and the Dirac value $\alpha = \frac{g-2}{2}$. For the electron, the anomalous moment $\alpha_{e} = \frac{(g-2)_e}{2}$ computed from Quantum Electrodynamics matches with the experimental value up to a striking precision of $10^{-12}$.

However recently, new experimental results from the Muon g-2 collaboration \cite{Abi2021} have confirmed a discrepancy between the observed value of the muon anomalous moment $\alpha_{\mu}$ and the standard model prediction \cite{Patrignani2016,Aoyama2020}. The combined data from the Brookhaven \cite{Bennett2004} and the Fermilab Muon g-2 \cite{Abi2021} experiments lead to a $4.2 \sigma$  discrepancy
\begin{equation}
 \Delta \alpha_{\mu} = \alpha_{\mu, EXP} - \alpha_{\mu, SM} = (251 \pm 59) \times 10^{-11}  .
\end{equation}
The discrepancy has fueled the discussion upon the need for new physics, since, if confirmed, it would be a clear indication of physics beyond the SM. Many speculations about the origin of this discrepancy have been made in the last years, including the contributions from axion-like particles \cite{Marciano2016} and dark photons (\cite{Cazzaniga2021} and refences therein), which, up to now, have not been experimentally detected.

On the other hand, nonlocal theories in which the field quantities are not restricted to be
always point functions in the ordinary space, have proliferated since the seminal work of Yukawa \cite{Yukawa}. Nonlocal theories of gravity have been developed as a extension of general relativity. Here nonlocal (i.e. non-polynomial) form factors in the gravitational action can help to solve the problem of ghosts as well as to improve the ultraviolet behavior of the quantized theory \cite{nonlocalG1,nonlocalG2,nonlocalG3,nonlocalG4,nonlocalG5,nonlocalG6,nonlocalG7,nonlocalG8,nonlocalG9}.
In the particle physics context, nonlocal theories are closely tied to string theories \cite{kato,Calcagni2014,Witten1986,Kostelecky1990,Kostelecky1988,Freund1987,Freund19872,Brekke1988,Frampton1988,Dragovich2007,Douglas2021,Gross1990,Brezin1990,Ghoshal2006}, as well as noncommutative field theories \cite{noncommloc1,noncommloc2}. Recently, a nonlocal extension of the standard model has been proposed \cite{Biswas}.

In this paper we speculate about the possibility that the $g-2$ discrepancy for the muon be induced by a nonlocal theory of the kind discussed in \cite{Biswas}. We compute the leading order and next to leading order nonlocal correction to the anomalous magnetic moment $\alpha_{NL}$, and we find that it depends in a simple fashion on the nonlocality scale $M_f$ and the fermion mass $m_f$ as $\alpha_{NL} \propto \frac{m_f^2}{M_f^2}$.  We then compare the known experimental anomaly $\Delta \alpha_{\mu}$ with the nonlocal prediction  $\alpha_{\mu, NL}$ in order to find a lower bound on the nonlocality scale. Our results indicate that the nonlocal theory might explain  the discrepancy observed in the muon moment, and then $\alpha_{\mu, NL}$ might represent an important contribution to $\Delta \alpha_{\mu}$.  Moreover the nonlocal contribution provides a very high discrepancy in the tau moment. 

The paper is structured as follows. In sec.II, we analyze the muon magnetic moment in a non-local theory, by computing the tree level correction (subsection A) and the next to leading order correction (sebsection B) in non-local electrodynamics, and then we provide lower bounds on the non-locality scale. In sec. III, we report conclusions and discussions.

\section{Muon magnetic moment in a non-local theory}

\begin{figure}[t]
\includegraphics[width=\columnwidth]{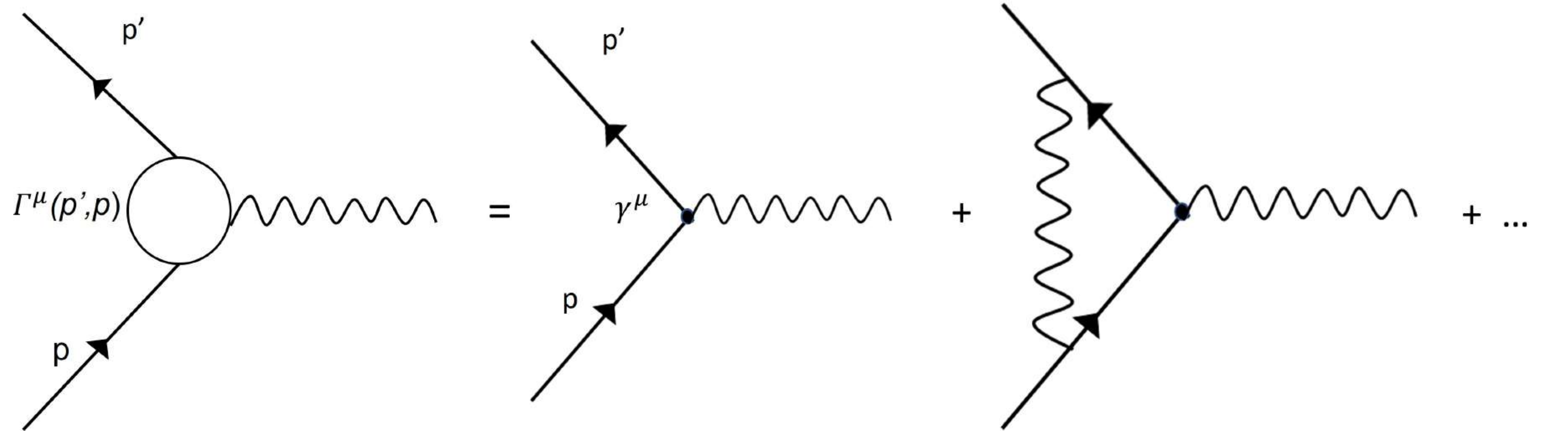}
\caption{Reference Feynman diagram for the calculation of the magnetic dipole moment. The vertex function $\Gamma^{\mu} (p',p)$ has to include all the relevant higher order diagrams (symbolized by the dots).}
\label{pdf_F}
\end{figure}

In deriving the magnetic moment of the charged leptons we follow the procedure and notation used in \cite{Peskin}, appropriately modified for the non-local theory described in \cite{Biswas}. We briefly outline how the magnetic moment is computed starting from Quantum Field Theory. The charged lepton field $\psi (x) $ is assumed to be coupled to a classical background electromagnetic field $A^{\mu}_{CL} (x)$ via the interaction Hamiltonian $H_I = e \int d^3 x A^{\mu}_{CL} (x) j_{\mu} (x)$, where $e$ is the electron charge and $j_{\mu} = \bar{\psi} (x) \gamma_{\mu} \psi (x)$ is the charged lepton current. The background field used to compute the magnetic moment is a static 3-vector field $A^{\mu}_{CL} (x) \equiv (0, \pmb{A}_{CL} (\pmb{x}))$ producing the magnetic field $\pmb{B} = \pmb{\nabla} \times \pmb{A}_{CL}$, with boldface letters denoting $3d$ spatial vectors. The amplitude for the scattering off the classical field $\pmb{A}_{CL} (\pmb{x})$ can be neatly expressed as
\begin{equation}\label{scatteringAmplitude}
 i \mathcal{M} = i e \left[ \bar{u} (p') \pmb{\Gamma} (p',p) u (p)\right] \pmb{\cdot} \pmb{A}_{CL} (\pmb{q}) \ .
\end{equation}
In the above equation $u(p)$ is the Dirac spinor corresponding to momentum $p$, $\pmb{q} = \pmb{p'}-\pmb{p}$ denotes the 3-momentum exchange, $\pmb{A}_{CL} (\pmb{q})$ is the Fourier transformed vector potential and $\pmb{\Gamma}(p',p)$ denotes the spatial part of the charged lepton vertex function.
The latter is computed perturbatively to any order in the fine structure constant $\alpha$ by including all the relevant diagrams (the reference diagram is shown in figure (1)) and can always be decomposed as
\begin{equation}\label{VertexFunctionDecomposition}
 \Gamma^{\mu} (p',p) = F_1 (q^2) \gamma^{\mu} + F_2 (q^2) \frac{i \sigma^{\mu \nu} q_{\nu}}{2 m} \ ,
\end{equation}
where $q=p'-p$ is the $4$-momentum exchange, $m$ is the charged lepton mass and $\sigma^{\mu \nu} = \frac{i}{2} [\gamma^{\mu},\gamma^{\nu}]$ denotes the commutator of Dirac gamma matrices. It is really the form factors ($F_1, F_2$) that one wishes to compute from the underlying theory, and it can be shown, by a straightforward argument, that they determine the $g$-factor ($\pmb{\mu}= g \left( \frac{e}{2m}\right) \pmb{S}$) via the simple relation
\begin{equation}\label{GFactor1}
 g = 2 \left( F_1 (q^2 \rightarrow 0) + F_2 (q^2 \rightarrow 0) \right) \ .
\end{equation}
The zeroth order g-factor in $\alpha$ coincides with the Dirac result $g_0 = 2$. Higher orders in perturbation theory, as well as the contributions due to electroweak and strong interactions bring the actual value of $g$ slightly above $2$ yielding the complete standard model prediction $g_{SM}$.

\subsection{Tree level correction in non-local electrodynamics}

\begin{figure}[t]
\includegraphics[width=0.6\columnwidth]{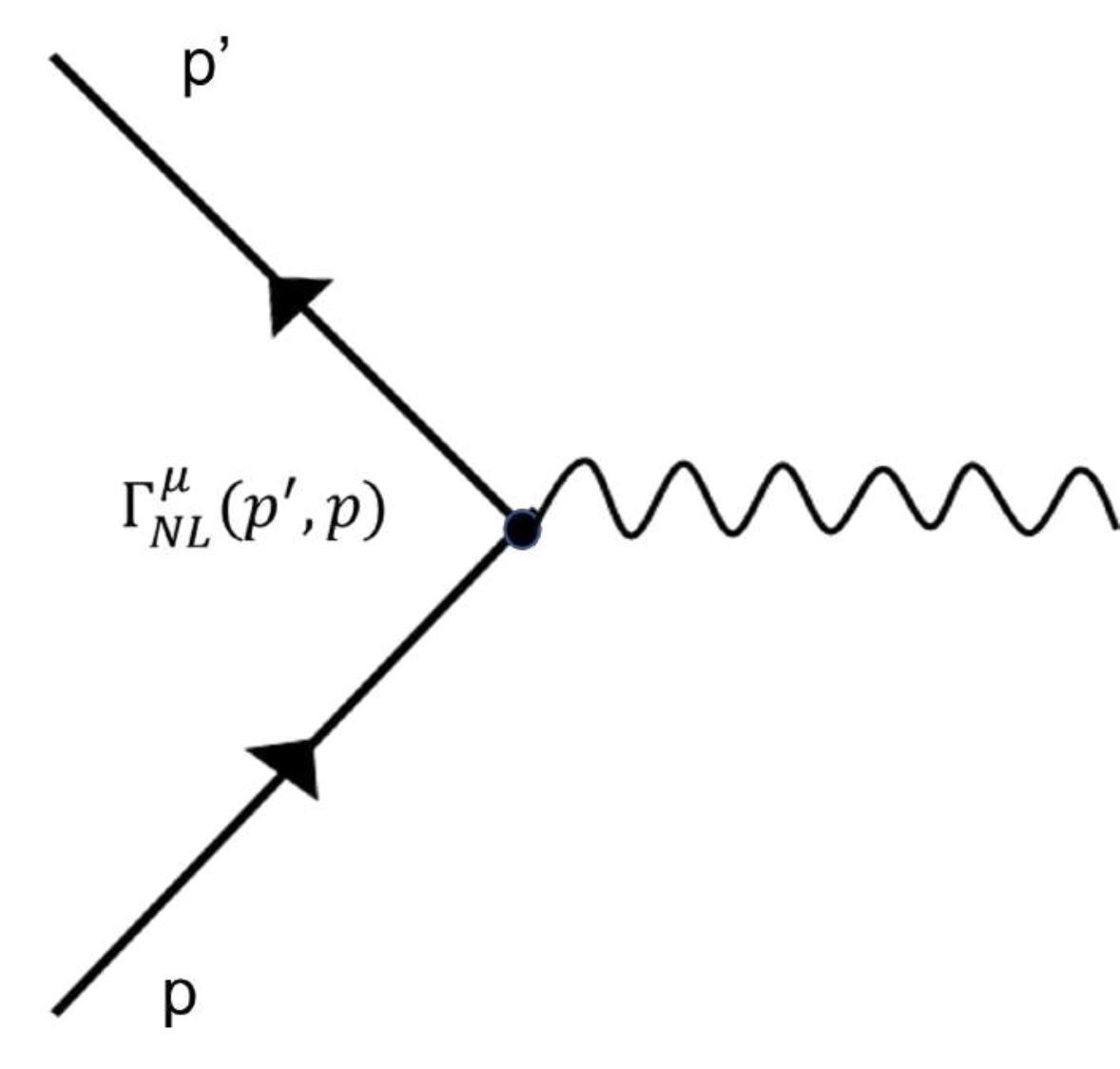}
\caption{Lowest order non-local Feynman diagram for the computation of the magnetic moment.}
\label{pdf_2}
\end{figure}

If standard electrodynamics is to be replaced with a non-local theory, the corrections to the magnetic moment are already present at tree level. To be precise, here we adopt the non-local theory proposed in \cite{Biswas}. The theory introduces the new mass scales $M_f$, the non-locality scale for the fermions, which may be either flavor-blind $M_e = M_{\mu} = M_{\tau} = M$ or flavor dependent $M_e \neq M_{\mu} \neq M_{\tau}$ , and $M_g$, the non-locality scale for the photons. The largest correction to the diagram of figure (1) due to non-locality comes from the lowest order ($\alpha^0$) non-local diagram of figure (2). Here one simply replaces the zeroth order local vertex function $\Gamma^{\mu} = \gamma^{\mu}$ with the zeroth order (in $\alpha$) non-local vertex function $\Gamma^{\mu}_{NL}$. Notice that at this order the vertex function is insensible to $M_g$, because no internal photon propagators appear. The non-local vertex function is found (apart from a $-ie$ factor) in \cite{Biswas}, and reads
\begin{widetext}
\begin{equation}\label{NonLocalVertex}
 \Gamma^{\mu}_{NL} (p',p) = \frac{1}{2} \left[ (p^{\mu} \slashed{p}' + p^{' \mu} \slashed{p})\left(\frac{e^{\frac{p^{'2}}{M_f^2}}-e^{\frac{p^{2}}{M_f^2}}}{p^{'2}-p^2}\right)+\left(e^{\frac{p^{'2}}{M_f^2}}+e^{\frac{p^{2}}{M_f^2}} \right) \gamma^{\mu}\right]  \ \ .
\end{equation}
The Feynman slash notation $\slashed{a} = \gamma^{\mu}a_{\mu}$ is used. In order to deduce the form factors, we must bring the equation \eqref{NonLocalVertex} into the form of equation \eqref{VertexFunctionDecomposition}. It clearly suffices to transform only the first term of equation \eqref{NonLocalVertex}. By sandwiching it between $\bar{u} (p')$ and $u(p)$ we find
\begin{eqnarray}\label{FirstVertexTerm}
 \nonumber && \bar{u}(p') \left[(p^{\mu} \slashed{p}' + p^{' \mu} \slashed{p})\left(\frac{e^{\frac{p^{'2}}{M_f^2}}-e^{\frac{p^{2}}{M_f^2}}}{p^{'2}-p^2}\right)  \right]u(p) = m\left(\frac{e^{\frac{p^{'2}}{M_f^2}}-e^{\frac{p^{2}}{M_f^2}}}{p^{'2}-p^2}\right) \bar{u}(p') (p^{' \mu} + p^{\mu}) u(p) \\
 && = m\left(\frac{e^{\frac{p^{'2}}{M_f^2}}-e^{\frac{p^{2}}{M_f^2}}}{p^{'2}-p^2}\right) \bar{u}(p')\left[2m \gamma^{\mu} - i \sigma^{\mu \nu} q_{\nu} \right]u(p) \ .
\end{eqnarray}
\end{widetext}
Here the first equality comes from the Dirac equation $\slashed{p} u(p) = m u(p)$ and the second from the Gordon equality \cite{Gordon}. We remark that the spinors $u(p)$ satisfy the same free Dirac equation as in the local theory. This is because the non-local action for the free ($A^{\mu}=0$) Dirac field \cite{Biswas} yields the Dirac equation unmodified. Indeed the fermion action for $A^{\mu}=0$ is \cite{Biswas}
\begin{eqnarray}
S_f = \int d^4 x \bar{\psi}_i e^{-\frac{\square}{M_f^2}} (i \slashed{\partial}-m_i)\psi_i
\end{eqnarray}
 and it is straightforward to see that upon variation it yields the ordinary Dirac equation for $\psi_i$. Recognizing the first term of equation \eqref{NonLocalVertex} from equation \eqref{FirstVertexTerm}, we immediately deduce the form factors
\begin{eqnarray}
 \nonumber F_1 &=& \frac{1}{2} \left[\left(e^{\frac{p^{'2}}{M_f^2}}+e^{\frac{p^{2}}{M_f^2}} \right) + 2m^2 \left(\frac{e^{\frac{p^{'2}}{M_f^2}}-e^{\frac{p^{2}}{M_f^2}}}{p^{'2}-p^2}\right) \right] \\
 F_2 &=& - m^2 \left(\frac{e^{\frac{p^{'2}}{M_f^2}}-e^{\frac{p^{2}}{M_f^2}}}{p^{'2}-p^2}\right)
\end{eqnarray}
or $F_1 + F_2 =\frac{1}{2} \left(e^{\frac{p^{'2}}{M_f^2}}+e^{\frac{p^{2}}{M_f^2}} \right) $.
It is clear that for external on-shell momenta $p^2 = m^2 = p^{'2}$, the form factors, and in particular $F_1$, depend explicitly on the fermion mass $m$. The nonlocal correction may be considered as giving rise to an effective flavor-dependent electric charge (compare eq. \eqref{NonLocalVertex} with the local tree level vertex $\Gamma^{\mu}_L = e \gamma^{\mu}$). This can be interpreted either as a violation of the lepton charge universality (implying, for instance, different muon and electron charges) or a violation of the electromagnetic gauge invariance and thus a violation of charge conservation.
In the spirit of the nonlocal theory proposed in \cite{Biswas}, preserving the electromagnetic gauge invariance, a violation of the lepton charge universality is required. A process like the muon decay $\mu^{-} \longrightarrow e^{-} + \bar{\nu}_e + \nu_{\mu}$ is consistent with electric charge conservation only if neutrinos are allowed an electric charge. This is possible in view of an eventual neutrino millicharge \cite{Giunti2015,Studenikin2014,Studenikin2020}. Therefore the possibility of a neutrino electric millicharge may justify a flavor-dependent electric charge.

Since the background field is static, no energy exchange occurs $q^0 = 0 = p^{'0}-p^{0}$ and the $q^2 \rightarrow 0$ limit implies $p^{'\mu} \rightarrow p^{\mu}$. On the other hand, the external fermion momenta are necessarily on-shell $p^2 = p'^2 = m^2$, so that from equation \eqref{GFactor1} we find the $g$-factor
\begin{eqnarray}\label{GFactor2}\nonumber
 g  &=&  2\left( F_1 (q^2 \rightarrow 0) + F_2 (q^2 \rightarrow 0) \right) = 2e^{\frac{m^2}{M_f^2}}
\\
 &\simeq & 2\left(1+\frac{m^2}{M_f^2} \right)
\end{eqnarray}
The last equivalence holds for $m \ll M_f$, as it is always the case.

\begin{figure}[b]
\includegraphics[width=0.6\columnwidth]{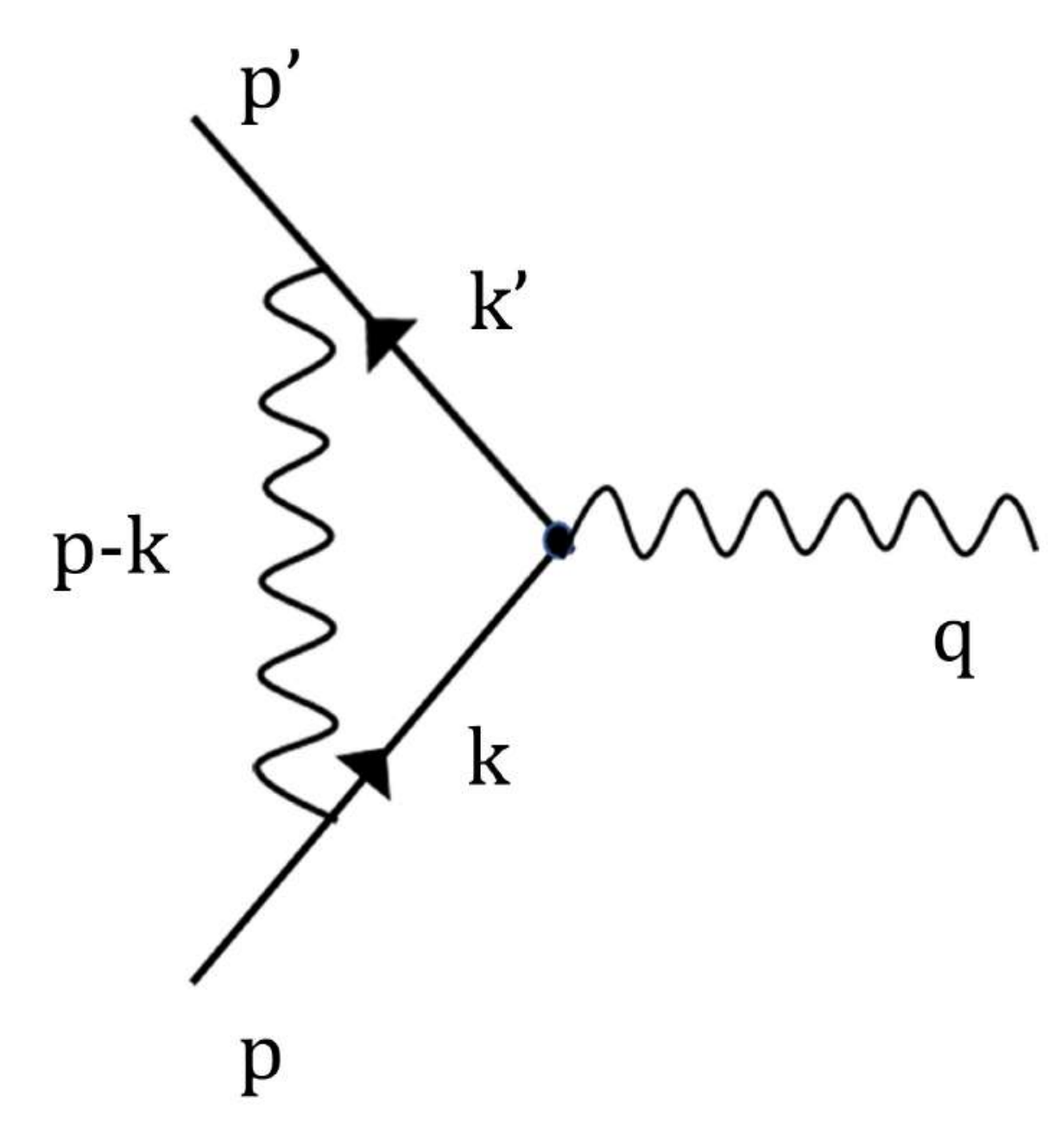}
\caption{Reference Feynman diagram for the order $\alpha$ contribution to the vertex function $\Gamma^{\mu}$.}
\label{pdf_u}
\end{figure}

A couple of comments are now in order. First, the tree-level non-local correction to the $g$-factor $(g-2)_{NL} = 2 \left(e^{\frac{m^2}{M_f^2}} - 1 \right) \simeq \frac{2m^2}{M_f^2}$
represents an additional contribution to the full (local) standard model prediction $(g-2)_{SM}$. Secondly, the dependence of the $g$-factor on the fermion mass $\propto m^2$ prefigures an interesting quantitative prediction. If $M_f$ is flavor-blind, we can immediately read off Eq. \eqref{GFactor2} that the muon correction $(g-2)_{\mu, NL}$ is about $40.000$ times larger than the correction for the electron, because $(g-2)_{\mu, NL} \simeq \frac{m_{\mu}^2}{m_e^2} (g-2)_{e, NL}$. For the same reason, the $\tau$ correction is about $\frac{m_{\tau}^2}{m_{\mu}^2} \simeq 282$ times larger than the muon correction. Therefore our computations allow to describe the observed anomalies for electron and muon magnetic momentum and to predict greater deviations from standard model prediction in the case of the tau magnetic momentum.

\begin{widetext}
\subsection{Next to leading order non-local correction}

The vertex function $\Gamma^{\mu}_{NL} (p',p)$ has, roughly speaking, two expansion parameters, the non-locality scale $M_f, M_g$ and the fine structure constant $\alpha$. Denoting by $(m,n)$ the expansion to order $m$ in $\frac{1}{M_{f,g}^2}$ and of order $n$ in $\alpha$, the tree level correction of the previous subsection corresponds to $(1,0)$. Here we wish to compute the next order in $\alpha$, namely the expansion to order $(1,1)$. We can immediately estimate that such a contribution shall be smaller by a factor $\alpha \simeq 1/137$ with respect to the leading order. In particular we consider the non-local version of the order $\alpha$ diagram of figure ({\ref{pdf_u}}).

We closely follow the computation of the local correction shown in \cite{Peskin}. In the non-local counterpart of diagram (\ref{pdf_u}) each of the inner elements (the three propagators and the three vertices) are replaced by their non-local form, namely

\begin{equation}
 \Pi_{\gamma,NL}^{\mu \nu} (p^2) = \frac{-i \eta^{\mu \nu}}{p^2 + i \epsilon} e^{-\frac{p^2}{M_g^2}} \ ; \ \ \ \ \Pi_{f,NL} (p^{\mu}) = \frac{i (\slashed{p} + m)}{p^2 - m^2 + i \epsilon} e^{-\frac{p^2}{M_f^2}} \ ,
\end{equation}

for the propagators and equation \eqref{NonLocalVertex} for the vertices. The correction we wish to compute is then, referring to figure (\ref{pdf_u}) 

\begin{equation}\label{Nonlocalvertex2}
 \delta \Gamma^{\mu} (p',p) = \int \frac{d^4 k}{(2\pi)^4} \Pi^{\gamma, NL}_{\nu \rho} ((k-p)^2) \bar{u} (p') (-i e \Gamma^{\nu}_{NL}(p',k')) \Pi_{f,NL}(k') \Gamma^{\mu}_{NL} (k',k) \Pi_{f,NL}(k) (-i e \Gamma^{\rho}_{NL}(k,p)) u(p) \ .
\end{equation}

Employing the Feynman parameters $x,y,z$ we can bring the denominator of the integrand to the form $\frac{2}{D^3}$, with $D= l^2 - \Delta + i \epsilon$, $\Delta = -xyq^2 + (1-z)^2 m^2$ and $l = k + yq -zp$. The integration momentum variable is switched to $l$ and a further integration on the Feynman parameters $\int_0^1 \ dx \ dy \ dz \ \delta (1-x-y-z) $ is introduced. Each of the non-local elements brings an exponential factor of the form $e^{\frac{\pm s^2}{M^2}}$ into the numerator of equation \eqref{Nonlocalvertex2}, where $s$ is some linear combination of the internal momentum variable $l$ and the external momentum variables $p,p',q$ and $M$ stands for either $M_g$ or $M_f$. Expanding the exponentials we get 

\begin{equation}
 e^{\pm s^2 / M^2} = 1 \pm \frac{l^2 + l\cdot(...) + (p_{ex})^2}{M^2} + \frac{1}{2 M^4} \left(l^4 + (...) + (p_{ex})^4 \right) \pm ...
\end{equation}

Where $p_{ex}$ denotes a combination of external momenta. 
The odd powers of $l$ do not contribute to the integral, since the denominator is even $D = D(l^2)$. The even powers of $l$, on the other hand, all lead to divergent expressions (logarithm and power law). When all the non-local terms are included, these divergences go well beyond the local logarithmic divergence arising from the local diagram \cite{Peskin} and therefore require a more sophisticated renormalization scheme to be reabsorbed, with respect to local electrodynamics. The full renormalization of the non-local theory is beyond the scope of this paper, and is demanded to future studies. Here we focus on the finite parts, arising from the terms of the form $\frac{(p_{ex})^{2n}}{M^{2n}}$. Since each of the external momenta is either on-shell ($p^2 = p^{'2} = m^2$) or shall be taken to vanish at the end of the calculation $q^2 \rightarrow 0$, it is sufficient to consider only the lowest orders $\frac{p_{ex}^2}{M^2}$. The next finite terms are indeed suppressed by higher powers of $m^2/ M^2$ (they are $O(\frac{m^4}{M^4})$).

\begin{figure}
\includegraphics[width=\columnwidth]{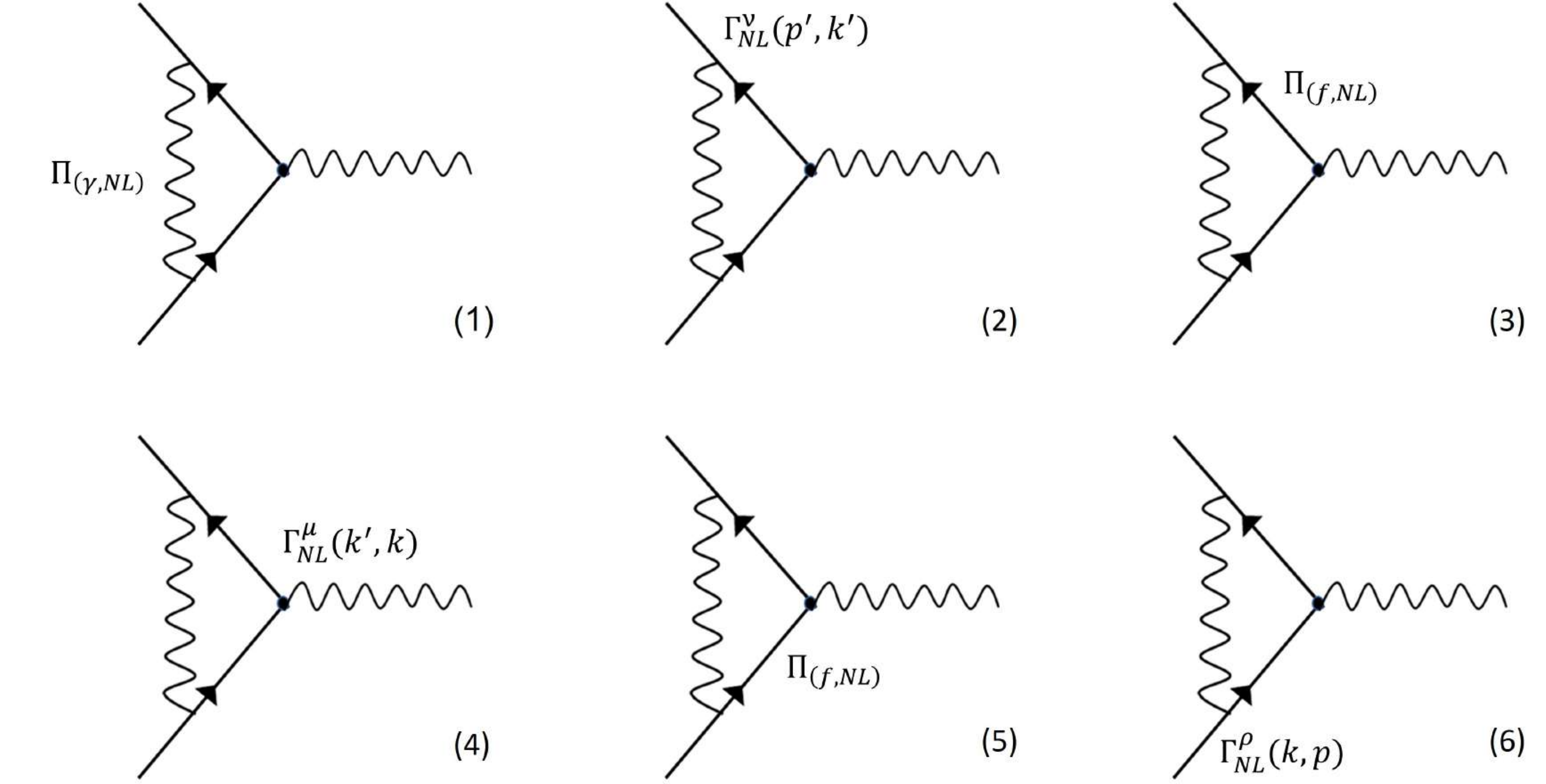}
\caption{Non-local vertex diagrams.}
\label{pdf_v}
\end{figure}

For a similar reason, to order $O (\frac{m^2}{M^2})$ it is sufficient to consider the diagrams in which only one of the inner elements is replaced by its non-local counterparts, as shown in the figure (\ref{pdf_v}). The order $O(1)$ contribution for the diagrams of figure (\ref{pdf_v}) is of course the same for all, and corresponds to the local order $\alpha$ correction to the vertex function. For instance, the diagram $(1)$ of figure (\ref{pdf_v}) corresponds to
\begin{eqnarray*}
 \delta \Gamma_1^{\mu} (p',p) &=& 2 i e^2 \int \ dx \ dy \ dz \ \delta (1-x-y-z) \int \frac{d^4 l}{(2 \pi)^4}\frac{2}{D^3} \left[ 1 - \frac{(k-p)^2}{M_g^2} \right] \bar{u}(p') \\ && \Bigg[\gamma^{\mu} \left(\frac{-l^2}{2} + (1-x)(1-y)q^2 + (1-4z +z^2)m^2 \right) 
 + i \frac{\sigma^{\mu \nu} q_{\nu}}{2m} \left(2m^2 z (1-z) \right)\Bigg] u(p)
\end{eqnarray*}

where it is understood that $(k-p)$ must be expressed in terms of $l$ and any term containing $l$ has to be discarded (either for symmetry or because it must be reabsorbed by the renormalization prescription). We distinguish two kinds of terms. The ``standard'' contributions have the general form
\begin{equation}\label{Standardcontribution}
 \delta \Gamma^{\mu, S} (p',p) = \int d \Omega f(p_{ex}^2)) \bar{u} (p')\Bigg[\gamma^{\mu} \left(\frac{-l^2}{2} + (1-x)(1-y)q^2 + (1-4z +z^2)m^2 \right) 
 + i \frac{\sigma^{\mu \nu} q_{\nu}}{2m} \left(2m^2 z (1-z) \right)\Bigg] u(p)
\end{equation}
where $d \Omega = \frac{
d^4l}{(2\pi)^4 D^3} \ dx \ dy \ dz \ \delta (1-x-y-z) $. These terms are essentially the same as the local integral, except for a function depending only on the external momenta $f(p_{ex}^2) \propto \frac{m^2}{M^2}$. After discarding the higher powers of $l$ and taking the $q^2 \rightarrow 0$ limit, the computation of these integrals is straightforward. All the propagator terms (diagrams $(1,3,5)$ of figure \ref{pdf_v}) as well as the bare $\gamma^{\mu}$ part (see equation \eqref{NonLocalVertex}) of the vertex diagrams ($(2,4,6)$ of figure \ref{pdf_v}) yield standard contributions. 
Let us work out the integral corresponding to the first diagram of figure \ref{pdf_v}. Discarding the local term and taking the $q^2 \rightarrow 0$ limit, the integral becomes

\begin{eqnarray}
\nonumber  \delta \Gamma_1^{\mu} (p',p) &=& - \frac{4i e^2}{M_g^2} \int_0^1 dx \ dy \ dz \ \delta (1-x-y-z) \int \frac{d^4 l}{(2\pi)^4} \frac{1}{(l^2 - \Delta)^3} \left( m^2 (1-z)^2\right) \\ 
&\times& \bar{u}(p') \left[ \gamma^{\mu} m^2 (1-4z + z^2) + \frac{i \sigma^{\mu \nu}q_{\nu}}{2m} \left(2m^2 z(1-z) \right) \right] u(p) \ .
\end{eqnarray}

Where it is understood that $\Delta = \Delta (q^2 \rightarrow 0)=m^2(1-z)^2$ and we have omitted the divergent terms, which can be removed by means of an appropriate regularization. The $l$ integration is most easily carried out by means of a Wick rotation, yielding

\begin{eqnarray}\label{FirstVertex}
 \nonumber  \delta \Gamma_1^{\mu} (p',p) &=& - \frac{\alpha}{\pi} \frac{m^2}{M_g^2} \int_0^1 \ dx \ dy \ dz \ \delta (1-x-y-z)  
 \bar{u}(p') \left[ \gamma^{\mu}  (1-4z + z^2) + \frac{i \sigma^{\mu \nu}q_{\nu}}{2m} \left(2 z(1-z) \right) \right] u(p) \ .
\end{eqnarray}

We can read off equation \eqref{FirstVertex} the corrections to the form factors, which after performing the integration over the Feynman parameters read $\delta F_1 = \frac{5\alpha}{3\pi} \frac{m^2}{M_g^2} $ and $\delta F_2 = -\delta F_1$. According to equation \eqref{GFactor1}, the first diagram gives no net contribution to the anomalous magnetic moment $\delta g = 2 (\delta F_1 + \delta F_2) = 0$. The other standard contributions are computed similarly and are collected in the table (I).

The vertex diagrams (namely $2,4,6$ of figure \ref{pdf_v}) also contain a part corresponding to the first term of equation \eqref{NonLocalVertex}. These terms cannot be immediately reduced to the form of equation \eqref{Standardcontribution}. For instance the ``non standard'' part of diagram $2$ is

\begin{equation}\label{NonStandardContribution}
 \delta\Gamma_2^{\mu, NS} = \frac{i e^2}{M_f^2} \int \frac{d^4 k}{(2 \pi)^4} \frac{\eta_{\nu \rho}}{(k-p)^2 + i \epsilon} \bar{u} (p') \left(k^{' \nu} \slashed{p}' + p^{' \nu} \slashed{k}' \right) \frac{i \left(\slashed{k}' + m \right)}{k^{'2}-m^2 + i \epsilon} \gamma^{\mu} \frac{i \left(\slashed{k}+m \right)}{k^2 - m^2 + i \epsilon} \gamma^{\rho} u(p) \ .
\end{equation}

Evaluating the correction to the form factors from integrals of the kind \eqref{NonStandardContribution} is a lengthy but simple process, and requires a massive use of the properties of the gamma matrices, along with the introduction of Feynman parameters. The results are collected in the table (I).
By summing all the corrections to the form factors, we finally find

\begin{equation}
 \delta g = 2 \left(\delta F_1 + \delta F_2 \right) = \frac{\alpha}{\pi} \frac{m^2}{M_f^2} \left(\frac{325}{12} -42 \ln 2 \right) \ .
\end{equation}

As expected this correction is about a factor $\alpha$ smaller than the tree level correction.

{\renewcommand{\arraystretch}{1.6}
\begin{table}\label{Table1} 
\begin{tabular}{||c | c | c ||}
 \hline
 Diagram & $\delta F_1$ & $\delta F_2$ \\ [0.5ex] 
 \hline\hline
 1 & $\frac{5\alpha}{3\pi} \frac{m^2}{M_g^2} $ & $-\delta F_1$  \\ 
 \hline
 2 (Standard part) & $\frac{\alpha}{4 \pi} \frac{m^2}{M_f^2} \left(\frac{31}{6}-16 \ln 2 \right)$ & $\frac{\alpha}{2 \pi} \frac{m^2}{M_f^2} \left(-\frac{19}{6}+4 \ln 2 \right)$ \\
 \hline
 3 & $\frac{\alpha}{\pi} \frac{m^2}{M_f^2} \left(\frac{23}{6}-12 \ln 2 \right)$ & $\frac{\alpha}{\pi} \frac{m^2}{M_f^2} \left(2-4 \ln 2 \right)$ \\
 \hline
 4 (Standard part) & $-\frac{\alpha}{\pi} \frac{m^2}{M_f^2}$ & $\frac{\alpha}{2\pi} \frac{m^2}{M_f^2} \left(-\frac{7}{6} + 2 \ln 2 \right)$  \\
 \hline
 5 & $\frac{\alpha}{\pi} \frac{m^2}{M_f^2} \left(\frac{23}{6}-12 \ln 2 \right)$ & $\frac{\alpha}{\pi} \frac{m^2}{M_f^2} \left(2-4 \ln 2 \right)$ \\
 \hline
 6 (Standard part) & $\frac{\alpha}{4 \pi} \frac{m^2}{M_f^2} \left(\frac{31}{6}-16 \ln 2 \right)$ & $\frac{\alpha}{2 \pi} \frac{m^2}{M_f^2} \left(-\frac{19}{6}+4 \ln 2 \right)$ \\
 \hline
 2 (Non standard) & $\frac{\alpha}{ \pi} \frac{m^2}{M_f^2} \left(\frac{47}{12}-10 \ln 2 \right)$ & $\frac{\alpha}{ \pi} \frac{m^2}{M_f^2} \left(-\frac{3}{2}+ 23 \ln 2 \right)$ \\
 \hline
 4 (Non standard) & $\delta F_1$ & $- \delta F_1$ \\
 \hline
 6 (Non standard) & $\frac{\alpha}{4 \pi} \frac{m^2}{M_f^2} \left(\frac{13}{2}-12 \ln 2 \right)$ & 0 \\ [1ex] 
 \hline
\end{tabular}
\caption{List of the corrections from the diagrams of figure \ref{pdf_v}.}
\end{table}}

\end{widetext}
\subsection{Lower bounds on the non-locality scale}

The non-local correction might explain, at least partially, the so-called muon $(g-2)$ anomaly, that is, the difference between the observed $\alpha_{\mu, EXP} = \frac{(g-2)_{\mu, EXP}}{2}$ and the standard model prediction $\alpha_{\mu, SM} = \frac{(g-2)_{\mu, SM}}{2}$. The observed anomalous magnetic moment of the electron $\alpha_{e, EXP}$ is instead much closer to the standard model prediction $\alpha_{e, SM}$. This fact is in line with the generic prediction of the non-local theory, which for a flavor-blind scale $M = M_{e} = M_{\mu} = M_{\tau}$ entails a larger anomaly for the muon (and for the tau).
Attributing the difference in the observed values of $\alpha$ and the standard model computation to the non-local correction, we now determine the lower bounds on $M_f$, for both flavor-blind and flavor-dependent scenarios.

In the flavor-blind case the relevant inequality is
\begin{eqnarray}\label{NonLocalScale}
 \frac{m_{\mu}^2}{M^2} \leq \alpha_{\mu , EXP} - \alpha_{\mu, SM} = 0.00000000257
\end{eqnarray}
which yields the lower bound $M \geq 4.384  \ \mathrm{TeV}$. Here we have used the latest available average values for the Standard Model prediction \cite{Patrignani2016} and the experimental result \cite{Abi2021} .The corresponding non-local correction to the electron anomaly is $\alpha_{e, NL} \leq 6 \times 10^{-14}$ which is well within the current experimental uncertainty $\sim 26 \times 10^{-14}$ \cite{PDG2020}. On the other hand, the mass dependence implies that for a flavor-blind scale, the corresponding maximum correction to the $\tau$ anomaly would be quite large $\alpha_{\tau, NL} \leq 7.267 \times 10^{-7}$.

In the flavor dependent case, the estimate for the muon scale is unaltered $M_{\mu} \geq 4.384 \mathrm{TeV}$. The electron scale should in any case be large enough that the non-local correction be smaller than the current experimental uncertainty on $\alpha_e$. This entails a slightly looser bound on $M_e$, namely $M_e \geq 1.004 \ \mathrm{TeV}$. In the end , the experimental uncertainty on the $\tau$ magnetic moment is still too high to deduce any plausible bound on $M_{\tau}$.

A comment is in order. The non-local scale, whose effect is to smear the point-like vertices of the Standard Model \cite{anish}, has been recently studied in \cite{paganis}. Here the authors show that a modification induced by non-locality to the heavy-boson cross section may enhance the production of boosted Higgs boson, providing a signal of non-local quantum field theory at the LHC. As pointed out in \cite{paganis}, the data of LHC 2 run will be sensitive to to non-local scale $\sim \text{few} \mathrm{TeV}$, which is of the same order of magnitude inferred for muon $g-2$ anomaly, as discussed in the present paper.

\section{Discussion and Conclusions}

In the framework of nonlocal theories, we have computed at tree level and at next to leading order the lepton magnetic  moments
 and  shown that the discrepancy between the observed value of the muon anomalous moment and the standard model prediction can be completely explained by simply replacing the zeroth order local vertex function $\gamma^{\mu}$ with the zeroth
order (in $\alpha$) non-local vertex function $\Gamma^{\mu}_{NL}$.

The correction derived for the magnetic momentum depends on the nonlocality mass scale $M_f$ introduced by the theory and on   the fermion mass $m_f$, i.e.:  $\alpha_{NL}\propto \frac{m_f^2}{M_f^2}$. The behavior of such correction allows in principle to explain the muon magnetic momentum anomaly if the nonlocality scale, according to equation \eqref{NonLocalScale}, is about $M \simeq 4.384 \mathrm{TeV}$. For larger $M$ the non-local theory is not sufficient to account for the observed discrepancy, and other mechanisms must come into play.   

In addition the form $\alpha_{NL}\propto \frac{m_f^2}{M_f^2}$ is compatible with the very small deviation of the electron magnetic moment from the standard model prediction, and allows to predict a larger magnetic momentum anomaly for the tau.
Finally, we have  determined the lower bounds on the nonlocality scale, for both flavor-blind and flavor-dependent scenarios.

 Our results are compatible with the ones derived in the local context, by including the contributions from axion-like particles   and dark photons.
 The mechanism here proposed could then contribute, together with other mechanisms beyond the Standard Model, to the explanation of the experimentally revealed anomalous magnetic moments of the charged leptons.

\section*{Acknowledgements}

A.C., G.L. and A.Q. acknowledge partial financial support from MIUR and INFN, A.C. and G.L. also acknowledge  the COST Action CA1511 Cosmology
and Astrophysics Network for Theoretical Advances and Training
Actions (CANTATA).

\end{document}